# What Causes Cost Overrun

# in Transport Infrastructure Projects?

By

Bent Flyvbjerg, Mette K. Skamris Holm and Søren L. Buhl

**Biographical notes**

**Bent Flyvbjerg** is a professor of planning at Aalborg University, Denmark. He is founder and director of the university's research program on large scale transport infrastructure planning. His latest books are *Rationality and Power* (University of Chicago Press, 1998) and *Making Social Science Matter* (Cambridge University Press, 2001). He is principal author of *Megaprojects and Risk* (Cambridge University Press, 2003). **Mette K. Skamris Holm** is an assistant professor of planning at Aalborg University and a research associate with the university's research program on large scale transport infrastructure planning. Her main interest is economic appraisal of projects. **Søren L. Buhl** is an associate professor of mathematics at Aalborg University. He is associate statistician with the university's research program on large scale transport infrastructure planning.

**Abstract**

This article presents results from the first statistically significant study of causes of cost escalation in transport infrastructure projects. The study is based on a sample of 258 rail, bridge, tunnel and road projects worth US$90 billion. The focus is on the dependence of cost escalation on (1) length of project implementation phase, (2) size of project and (3) type of project ownership. First, it is found with very high statistical significance that cost escalation is strongly dependent on length of implementation phase. The policy implications are clear: Decision makers and planners should be highly concerned about delays and long implementation phases because they translate into risks of substantial cost escalations. Second, it is found that projects have grown larger over time and that for bridges and tunnels larger projects have larger percentage cost escalations. Finally, by comparing cost escalation for three types of project ownership--private, state-owned enterprise and other public ownership--it is shown that the oft-seen claim that public ownership is problematic and private ownership effective in curbing cost escalation is an oversimplification. Type of accountability appears to matter more to cost escalation than type of ownership.

## Cost escalation and its causes

On the basis of the first statistically significant study of cost escalation in transport infrastructure projects, in a previous article we showed that cost escalation is a pervasive phenomenon in transport infrastructure projects across project types, geographical location and historical period (Flyvbjerg, Holm and Buhl, forthcoming). More specifically we showed the following (all conclusions highly significant and most likely conservative):

- Nine out of ten transport infrastructure projects fall victim to cost escalation (N=258).

- For rail average cost escalation is 45% (N=58, sd=38).

- For fixed links (bridges and tunnels) average cost escalation is 34% (N=33, sd=62).

- For roads average cost escalation is 20% (N=167, sd=30).

- For all project types average cost escalation is 28% (N=258, sd=39).

- Cost escalation exists across 20 nations and five continents; it appears to be a global phenomenon (N=258).

- Cost escalation appears to be more pronounced in developing nations than in North America and Europe (N=58, data for rail only).

- Cost escalation has not decreased over the past 70 years. No learning seems to take place (N=111/246).

The sample used to arrive at these results is the largest of its kind, covering 258 transport infrastructure projects in 20 nations worth approximately US$90 billion (1995 prices). In the present article we use this sample to analyse causes of cost escalation in transport infrastructure projects. By 'cause' we mean 'to result in'; the cause is not the explanation of the result. The main purpose of this article has been to identify which factors cause the cost escalation, to a lesser degree the reasons behind why they cause it. We test how cost escalation is affected by three variables: (1) length of implementation phase measured in years, (2) size of project measured in costs and (3) three types of ownership including public and private. In addition, we test whether projects grow larger over time. Results from a separate study of political explanations of cost escalation have been published in Flyvbjerg, Holm and Buhl (2002).

For all 258 projects in the sample we have data on percentage cost overrun. When we combine percentage cost overrun with other variables, for instance length of implementation phase, the number of projects becomes lower because data on other variables is not available for all 258 projects. For each added variable, we mention below for how many projects of the 258 data is available. As far as possible, all projects are used in each analysis. In no case have we omitted available data, except for the mentioned cases of outliers. Ordinary analysis of variance and regression analysis have been used for analysing the data. When talking about significance below, we use the conventional terms: very strong significance ($p < 0.001$), strong significance ($0.001 \leq p < 0.01$), significant ($0.01 \leq p < 0.05$), nearly significant ($0.05 \leq p < 0.1$) and non-significant ($0.1 \leq p$).

The present article is a companion paper to our article "How common and how large are cost overruns in transport infrastructure projects?", published in TRV 23(1), pp. 71-88, 2003. For a full description of the sample, data collection and methodology, we refer readers to the previously published article.

## Are sluggish projects more expensive?

The Commission of the European Union recently observed that the 'inherent sluggishness' of the preparation, planning, authorisation and evaluation procedures for large infrastructure projects creates obstacles to the implementation of such projects (Commission of the European Union 1993: 76). There is a fear that obstacles in the planning and implementation phases translate into cost escalation, if they do not block projects altogether (Ardity, Akan and Gurdamar 1985, Morris and Hough 1987, Snow and Dinesen 1994, Chan and Kumaraswamy 1997).

We decided to test whether such fear is corroborated by the empirical evidence. More specifically, we decided to test the thesis that projects with longer implementation phases tend to have larger cost escalations. We here define length of implementation phase as is common, i.e. as the time period from decision to build until construction is completed and operations have begun. Cost development is defined as the difference between actual and forecast construction costs in percentage of forecast construction costs.

Information about length of implementation phase is available for 111 of the 258 rail, fixed link (bridges and tunnels) and road projects for which we have data on cost development (38 out of 58 rail, 33 of 33 fixed link, 40 of 167 road projects). Figure 1 shows the dependence of cost escalation on length of implementation phase. The figure suggests that there is a statistical relationship between length of implementation phase and cost escalation where a longer implementation phase tends to result in a larger cost escalation. Statistical tests corroborate this impression. The tests have been carried out with and without projects with implementation phases of 13 years and longer. The reason for the 13 year cut-off is that the assumptions for the regression analysis do not seem to be fulfilled for projects of longer duration, mainly linearity and homoscedasticity. Projects with implementation phases of 13 years and longer can be considered statistical outliers. This is revealed by residual plots and is most obvious for bridges and tunnels. For uniformity, the cut-off has been done for all groups. When the outliers are included, the results of analyses are less sharp, due to higher statistical error.

[Figure 1 app. here]

For the 101 projects with implementation phases known to be less than 13 years we find a highly significant dependence of cost escalation on length of implementation phase ($p<0.001$, t-test). The null hypothesis that length of implementation phase has no effect on cost escalation is falsified. Longer implementation phases significantly tend to translate into larger percentage cost escalations. The influence of length of implementation phase on cost escalation is not statistically different for rail, fixed link (bridges and tunnels) and road projects, respectively (p=0.159). We have chosen to treat the three types of projects on aggregate. Three regression lines could be given, one for each project type. However, the null hypothesis of a common regression line is in conformity with the data and gives a simpler model. The p-value is low but not close to 0.05. The regression line for cost escalation as a function of length of implementation phase is shown in Figure 1.

The equation for the regression line is:

$$\Delta C = 0.4 + 4.64 * T$$

where

$\Delta C$ = Cost escalation in % (fixed prices)

T = Length of implementation phase in years.

The detailed statistics are:

Intercept: mean = 0.448, sd=8.258, t=0.054, p=0.957.

Slope: mean= 4.636, sd=1.279, t=3.626, p=0.00048.

R-square = 0.1172.

The 95% confidence interval for the slope is 2.10 to 7.17. The confidence interval gives the uncertainty of the analysis. It is of course important that zero is not included in the interval.

Given the available evidence, we see that for every passing year from the decision to build a project until construction ends and operations begin, we must expect the project to incur an average increase in cost escalation of 4.64 percentage points. Thus for a 1 billion dollar project, each year of delay would cost on average 46 million dollars. For a project in the size-range of the Channel tunnel between France and the UK, the expected average cost of delay would be approximately 350 million dollars per year, or about a million dollars per day.

We note that these figures include only construction costs, i.e. not financing costs. With financing costs included the figures would be considerably higher and would be even more sensitive to the time factor, because financing costs consist mainly of accrued interests. Financing costs are particularly sensitive to long delays, because delays defer income while interest, and interest of interest, keep accumulating. Long delays may result in projects ending up in the so-called 'interest trap', where a combination of escalating construction costs, delays and increasing interest payments result in a situation where income from a project cannot cover costs. This has happened, for instance, for the two longest underwater rail tunnels in Europe, the Channel tunnel and the Danish Great Belt rail link, which both had to be financially reorganised. The Øresund link between Sweden and Denmark has also run into problems of this kind, but it is too early in the life of this project, which opened in 2000, to say whether the result will be financial non-viability (Flyvbjerg, Bruzelius and Rothengatter 2003).

The average length of implementation phase is significantly different for different types of projects (p=0.002, F-test). Figure 2 shows a box plot for type of project and length of implementation phase. Fixed link projects (bridges and tunnels) have the longest implementation phase with an average of 6.6 years (sd=3.4), followed by rail projects with 6.3 years (sd=3.3) and roads with 4.3 years (sd=2.2). Consequently, cost escalation must be expected to be different for the three types of projects, and especially for road projects compared with rail and fixed link projects, because the length of implementation phases are different.

When considering the possibility of third factor or omitted variable effects on the results, one might speculate that the complexity of projects may be of importance to the size of cost escalations, i.e. some projects turn out to be more complex and this may result in larger cost escalations for such projects. Complexity is difficult to operationalise for statistical analysis, but the sample does not seem to include a bias concerning complexity. Thus the results appear to reflect differences between projects regarding length of implementation phase and not regarding complexity. Further investigations of complexity could be interesting but would involve other methods of analysis than those employed here.

[Figure 2 app. here]

In sum, excluding the most sluggish projects, i.e. those with an implementation phase of 13 years and longer, there is no statistical evidence that group of project has influence on cost escalation besides what can be explained by sluggishness. Length of implementation phase is the essential predictor and, as long as more evidence has not been found, it must be considered a stand-alone. Knowing length of implementation phase, we do not need to distinguish between rail, fixed link and road projects. It should be mentioned, however, that this conclusion is based on only the 111 projects for which information on length of implementation phase was available out of the 258 projects in the complete sample. Further, for the most sluggish projects the data do not allow firm conclusions. If we do not know length of implementation phase and only the project type is given, then road projects would have less cost escalation than fixed link projects. The important result to note here, however, is that if information on implementation duration is given, project type is not important.

Introducing into the analysis the geographical location of projects--in Europe, North America and other geographical areas, respectively--we find, firstly, that the influence of geographical location on length of implementation phase (cost escalation not considered) was very strong and statistically significant for fixed links and roads, with North America showing shorter implementation phases than other geographical areas ($p<0.001$). For rail there was no significant relationship. Secondly, we find that if length of implementation phase and geographical location are both known, then the same regression lines can be used for the three types of geographical location, with the proviso that only rail projects are included in our study for 'other geographical areas'.

The regression lines can be assumed to be parallel (see below for an explanation of why the slope for all projects above is different from the slope of the parallel lines for geographical areas):

Europe: $\Delta C = 14.2 + 3.28*T$

North America: $\Delta C = -1.3 + 3.28*T$

Other geographical areas: $\Delta C = 56.2 + 3.28*T$

where

$\Delta C$ = Cost escalation in % (fixed prices)

T = Length of implementation phase in years.

The 95% confidence interval for the slope is 0.58 to 5.97. The p-value for parallellity is p=0.967 Whereas the deviation of intercept for other geographical areas is significant, the difference in intercept between Europe and North America is only close to being significant (p=0.077). Further research is needed on this point. Logarithmic relationships were considered but rejected.

One may wonder why the slope is lower for the geographically subdivided data than for the undivided data. It is easy to see why this must be the case by conceiving three parallel 'clouds' of data points, one for each of the three geographical regions. Drawing one common regression line for all data points necessarily results in a slope higher than that of the regression lines for each individual 'cloud'. The observant reader may also observe that when considering to build a specific project, decision makers typically know in which geographical area the project would be located and that, therefore, the slope of 3.28 is more relevant in this case than the average slope for the whole dataset of 4.64.

In conclusion, the dependence of cost escalation on length of implementation phase is firmly established for transport infrastructure projects. We conclude, therefore, that there is good reason to be concerned about sluggish planning and implementation of such projects. Sluggishness may, quite simply, be extremely expensive. Consequently, before a project owner decides to go ahead and build a project, every effort should be made to conduct preparation, planning, authorisation and ex ante evaluation in a manner where such problems are negotiated and eliminated that may otherwise resurface as delays during implementation. Flyvbjerg, Bruzelius and Rothengatter (2003) describe ways in which this may be achieved. Similarly, after the decision to build a project, it is of crucial importance that the project organisation and project management are set up and operated in ways that minimise the risk of delays. If those responsible for a project fail to take such precautions aimed at pro-actively preventing delays and long implementation phases, the evidence indicate that the financiers--be they tax payers or private investors--are likely to be severely penalised in terms of cost escalations of a magnitude that could threaten project viability.

## Do bigger projects have bigger cost escalations?

Based on the results of the previous section, one might speculate that larger projects would have larger percentage cost escalations than smaller projects, because, other things being equal, implementation phases would be longer for larger projects with resulting increases in cost escalation. The question is, in short, whether larger projects are sluggish projects and therefore more prone to cost escalation?

Both the research literature and the media occasionally claim that the track record is poorer for larger projects than for smaller ones and that cost escalations for large projects are particularly common and especially large (Merewitz 1973: 278; Ellis 1985; Morris and Hough 1987: 1, 7). Until now it has been difficult or impossible to rigorously test such claims because data that would allow tests have been unavailable or wanting.

With the new and larger sample of data collected for the research reported in this article, we therefore decided to test whether cost escalation significantly vary with size of project. Forecast and not actual construction costs should be used here as measure of size of project for the following two reasons. First, cost escalation is statistically confounded with actual construction costs, being part of it, whereas forecast construction costs are not. Second, the decision regarding whether to go ahead with a given project is based on forecast construction costs; this is the decision variable, not actual costs.

As mentioned, we have percentage cost overrun for 258 projects. If we ask for the additional information, how is percentage cost overrun made up of forecasted and actual costs, this info is available for 131 of the 258 projects. Figure 3 shows the plot of percentage cost escalation against project size, with indication of project type, for these 131 projects. The plot shows no immediate dependence between the two variables. It also does not substantiate any thesis of different variability for smaller and larger projects. Analysis of covariance indicate that project types should be treated separately. Dummy variables could be used but are more error-prone in interpretation than the analysis presented below.

Tests done separately for rail, fixed link and road projects, show a nearly significant relationship between cost escalation and project size for fixed link projects (p=0.085), whereas there is no indication that percentage cost escalation depends on project size for road and rail projects (p=0.330 and p=0.496, respectively). If we refine the analyses further by again treating as statistical outliers projects with implementation phases of 13 years and longer, then percentage cost escalation significantly depends on project size for fixed links, with larger fixed links having larger percentage cost escalations (p=0.022). The regression line for fixed links without two statistical outliers is:

$$\Delta C \; = \; -28.9 + 23.0 * \log 10 (C_0)$$

where

$$\Delta C \; = \; \text{Cost escalation in \% (fixed prices)}$$

$C_0$ = Forecast costs of project (1995 level euros)

We conclude that for bridges and tunnels the available data support the claim that bigger projects have bigger percentage cost escalations, whereas this appear not to be the case for road and rail projects. For all project types, bigger projects do not have a larger risk of cost escalation than do smaller ones; the risk of cost escalation is high for all project sizes and types. We also conclude that the divisibility argument--that road and rail projects may have lower percentage cost overruns because they often can be phased in, whilst bridges and tunnels are only available once completed--is not supported by the data. Generally, the road projects are smallest. For fixed link and rail projects, Figure 3 indicates that the difference (between fixed link and rail projects) is significant also for large projects. The mega fixed link projects (actually the Channel tunnel and Great Belt bridge) do not have exceptional percentage cost overruns, a conclusion which runs counter to the divisibility argument. Finally, it should be mentioned that tests of correlation between project size and length of implementation phase show no significant results.

[Figure 3 app. here]

## Do projects grow larger over time?

Project size matters to cost escalation, we found above for bridges and tunnels. But even for projects where increased size correlate with neither bigger percentage cost escalations nor larger risks of escalation, as we found for rail and road projects, we wish to point out that there may be good practical reasons to pay more attention to--and use more resources to prevent--cost escalation in larger projects than in smaller ones. For instance, a cost escalation of, say, 50% in a 5 billion dollar project would typically cause more problems in terms of budgetary, fiscal, administrative and political dilemmas than would the same percentage escalation in a project costing, say, 5 million dollars. If project promoters and owners wish to avoid such problems, special attention must be paid to cost escalation for larger projects.

On this background we analysed the size of projects over time. Figure 4 shows the costs of projects plotted against year of completion. The figure is based on actual costs in order to show the real, as opposed to the budgeted, size of projects. Actual costs correspond to year of completion, which are also shown in the figure.

[Figure 4 app. here]

Correlation between time and cost is not immediately clear from figure 4. On closer statistical analyses, however, it turns out there is a significant increase over time in the size of road projects. The visual appearance of the data in the figure is rather different for the different types of projects, calling for different types of statistical analysis. Rail and road projects cluster in two groups according to year of completion, the road projects more distinctly. We have applied both a regression analysis and a two-sample comparison for these projects. For road projects, the regression line is (corrected for a statistical outlier):

$$\log 10(C_1) = 1.230 + 0.0098*(T - 1970)$$

where

$C_1$ = Actual costs of project (1995 level euros)

$T$ = Year of completion of project

corresponding to a 2.3% rise in project size each year, equivalent to a doubling in size in 30.8 years. The rise is statistically significant (p=0.011). There are two clusters of road projects with time spans 1954-1964 and 1987-1996. Using a two-sample comparison we find a significant increase in project size of 82.6% over the 32 years between the two clusters, corresponding to an annual increase of 1.9% (p=0.034, Welch's t-test).

For rail projects, the regression line is:

$$\log 10(C_1) = 2.43 + 0.0060*(T - 1970)$$

corresponding to an annual increase in project size of 1.4%. But now the rise is nonsignificant (p=0.582). Welch's two-sample test also produces a nonsignificant result.

For fixed links, the regression line is:

$$\log_{10}(C_1) \ = \ 2.322 + 0.0083*(T - 1970)$$

corresponding to a 1.9% rise in size each year. The result is non-significant, however (p=0.131). Two-sample testing is not suitable here.

Given the available evidence, we conclude that projects are growing larger over time, but only for road projects is such growth statistically significant. This may be explained by the fact that bridges, tunnels and rail projects tend to be larger and less divisible than road projects. Thus rail and fixed link projects have been large all along for the period under study and therefore have less scope for high percentage increases in size than road projects.

Granted the fact that project size is increasing, and granted that the same percentage cost escalation will typically cause more havoc in terms of budgetary, fiscal, administrative and political dilemmas in a large project than in a small one, we conclude that, other things being equal, an increase in project size translates into an increase in potential trouble for infrastructure development. For instance, a doubling in project size results in a doubling in additional fiscal demands for the same percentage cost escalation.

This, finally, translates into a need for an improved planning process and a better institutional set-up for infrastructure development and management, to prevent potential trouble from becoming real. For suggestions on how the planning process and institutional set-up for infrastructure development and management may be improved, see Bruzelius, Flyvbjerg and Rothengatter (1998) and Flyvbjerg, Bruzelius and Rothengatter (2003).

## Do private projects perform better than public ones?

During the past ten to twenty years, there has been a resurgence of interest in private sector involvement in the provision of infrastructure (Wright 1994; Seidenstat 1996; Flyvbjerg, Bruzelius and Rothengatter 2003: ch. 6). One main motive for this development has been a desire to tap new resources of funds to supplement the constrained resources of the public sector. Another central motive has been a widespread belief that the private sector is inherently more efficient than the public sector (Gómez-Ibáñez and Meyer 1993: 3-4; Ascher 1987; Moran and Prosser, 1994; Bailey and Pack 1995; Clark, Heilman and Johnson 1995-96).

Large cost escalations are typically seen as signs of inefficiency and in the research literature such escalations are often associated with public sector projects. One recent study speaks of 'the calamitous history of previous cost escalations of very large projects in the public sector' (Snow and Dinesen 1994: 172, Hanke 1987, Preston 1996, Gilmour and Jensen 1998). The study goes on to conclude that the 'disciplines of the private sector' can 'undoubtedly' play a large part in restraining cost escalations. Unfortunately, little evidence is presented here or elsewhere in the literature that would demonstrate that private projects do indeed perform better than public ones as regards cost escalation (Moe 1987, Bozeman 1988, Kamerman and Kahn 1989, Handler 1996).

Moreover, the evidence from what was intended as the international model of private financing, the Channel tunnel between France and the UK, actually points in the opposite direction with a cost escalation of 80%, or more than twice the average cost escalation of tunnels and bridges.

On this background we decided to test whether cost development varies with type of ownership of projects. Instead of using the conventional dichotomy public-private, we decided to operate with a slightly more complex trichotomy employing the following categories:

1. Private.

2. State-owned enterprise.

3. Other public ownership.

State-owned enterprises are corporations owned by government and are typically organised according to a companies act, for instance as incorporated or limited companies. Other public ownership is the conventional form of public ownership, with a ministry typically owning the project, which appears in the public budgets. Many variants of private and public and joint funding exist, with all sorts of conditions placed by lenders regarding interest rates, issues of risk and return, and packaging of project funding. However, with the available data, the grouping must necessarily be coarse to have enough data in each group for statistical analysis. A more detailed typology than that suggested above would be desirable at a later stage but is currently not possible because of lack of data to support it.

Our reasons for subdividing public projects into two different categories were grounded in results from previous research (Flyvbjerg, Bruzelius and Rothengatter 2003). Here we found that projects run by state-owned enterprises were subject to regulatory regimes that are significantly different from those found for projects under other public ownership. We concluded that such differences in regulatory regimes may influence performance differently.

More specifically, in research on the state-owned enterprises running the Great Belt and Øresund links--both multi-billion dollar projects linking Scandinavia with continental Europe--we had found that these projects may be subject to what we call the 'two stools' effect (Flyvbjerg, Bruzelius and Rothengatter 2003, ch. 7). The projects lack the transparency and public control that placement in the public sector proper would entail. On the other hand, we also found that the projects lack the competition and pressure on performance that placement in the private sector would bring about. In short, as regards accountability and performance, the Great Belt and Øresund projects might be said to 'fall between two stools', as the proverb has it. Following this line of reasoning, a recent report from the Danish Ministry of Finance singles out the Great Belt and Øresund projects as liable to a 'risk of lack of efficiency' during construction and operation due to 'lack of sufficient market pressure' (Finansministeriet 1993: 82).

However, our studies of the Great Belt and Øresund projects were basically two single-case studies. As such they did not permit statistically valid conclusions regarding the effects of ownership on performance. Now, with our sample of 258 transport infrastructure projects we wanted to see if the additional data would allow us to establish a more general pattern regarding ownership and performance.

We were able to establish ownership for 183 of the 258 projects in the sample. Again means and standard deviations dictate that we treat the three types of project separately in the statistical analyses. For fixed links, all types of ownership are represented, although sparsely (see Table 1). Tests for interaction with other explanatory variables indicate that ownership can be considered alone. Using a standard one-way analysis of variance, the effect of ownership on cost escalation is significant for fixed links (p=0.028). Looking at the means an interesting pattern emerges (see Table 1). State-owned enterprises show the poorest performance with an average cost escalation of 110%. Privately owned fixed links have an average cost escalation of 34%. Finally, and perhaps surprisingly, other public ownership shows the best performance with an average cost escalation of 'only' 23%.

[Table 1 app. here]

A test of whether the differences are due to differences between bridges and tunnels indicates that this is not the case, but the data are too few for firm conclusions. For 'other public' ownership against private ownership a classical non-paired t-test can be applied, with p=0.589. Therefore, although the mean for other public ownership is lower than for private ownership for fixed links this could be due to chance. We have also tested private and other public ownership as one group against state-owned enterprises. Pooling other public and private ownership may seem unusual, but it is substantiated by the data. With Welch's modification of the t-test we get that p=0.176, i.e. non-significance. Other public versus state-owned enterprise gives no significance either, with p=0.162.

The analyses of variance indicate significant differences in cost escalation for fixed links on account of ownership, but these differences cannot, at this stage, be located more precisely. Again we must conclude that even though our sample is relatively large when compared to other samples in this area of research, it is not large enough to support a subdivision into three types of projects combined with three types of ownership and still support firm statistical analysis. Further research should be done here with data for more fixed links.

Despite these reservations, one conclusion is clear from our analysis of ownership and cost development for fixed links: In planning and decision making for this type of project, the conventional wisdom, which holds that public ownership is problematic whereas private ownership is a main source of efficiency in curbing cost escalation, is dubious. This, of course, does not rule out the possibility that other reasons may exist for preferring private over public ownership; for instance, that private ownership may help protect the ordinary taxpayer from financial risk and may reduce the number of people exposed to such risk. But our study shows that the issue of ownership is more complex than usually assumed. We find that the problem in relation to cost escalation may not primarily be public versus private ownership. The problem appears more likely to be a certain kind of public ownership, namely ownership by state-owned enterprises. We expect further research on this issue to be particularly rewarding in either falsifying or confirming this finding.

For rail projects, private ownership is non-existent in our data. We therefore have only the dichotomy state-owned enterprise versus other public ownership. Table 2 shows average cost escalation for rail. For high-speed rail we again see that projects owned by state-owned enterprises have by far the largest cost escalation. The difference is highly significant (p=0.001, Welch t-test), but given the available data, which are scant and from projects on different continents, it is impossible to say whether the difference can be attributed to ownership alone or whether the geographical location of projects also play a significant role in affecting cost escalation. For instance, three Japanese, state-owned high-speed rail projects significantly influence the results and at this stage the data do not allow a decision as to whether this influence should be attributed to type of ownership or to the fact that the projects are Japanese, because ownership and geographical location are statistically confounded. For urban rail projects we find that state-owned enterprises perform better than 'other public' ownership, but this difference is non-significant (p=0.179). We conclude that for rail projects, too, further research is needed and can be expected to produce interesting results.

[Table 2 app. here]

Since all road projects in the sample fall in the category 'other public ownership' no analysis of the influence of ownership on cost escalation can be carried out here. This, again, is an area for further research, where data on privately owned roads and roads owned by state-owned enterprises can be expected to make a particularly important contribution.

## Conclusions

In a previous article we showed that large construction cost escalations in transport infrastructure projects are common and exist across different project types, different continents and different historical periods (Flyvbjerg, Holm and Buhl, forthcoming). In this article we test what causes construction cost escalation, focusing on three variables: (1) length of implementation phase, (2) size of project and (3) type of ownership. The database used in the tests are by no means perfect. A more robust database, with more, and more evenly distributed, observations across subdivisions, is desirable. Such a database is not available at present, however. The database provided is the best and largest that exists, and it is a major step ahead compared to earlier databases.

First, for length of implementation phase the main findings are:

- Cost escalation is highly dependent on length of project implementation phase and at a very high level of statistical significance ($p<0.001$).

- The influence of length of implementation phase on cost escalation is not statistically different for rail, fixed-link (bridge and tunnel) and road projects, respectively.

- For every passing year from the decision to build until operations begin, the average increase in cost escalation is 4.64 percentage points. For a project in the size-range of the Channel tunnel this is equal to an expected average cost of delay of approximately a million dollars per day, not including financing costs.

We conclude that decision makers should be concerned about long implementation phases and sluggish planning and implementation of large transport infrastructure projects. Sluggishness may, quite simply, be extremely expensive. Consequently, before a project owner decides to go ahead and build a project, every effort should be made to conduct preparation, planning, authorisation and ex ante evaluation in such ways that problems are negotiated and eliminated which may otherwise resurface as delays during implementation. Similarly, after the decision to build a project, it is of crucial importance that the project organisation and project management are set up and operated in ways that minimise the risk of delays. If those responsible for a project fail to do this, the evidence indicate that the financiers--be they tax payers or private investors--are likely to be severely penalised in terms of cost escalations of a magnitude that could threaten project viability.

Second, for size of project we find:

- For bridges and tunnels, larger projects have larger percentage cost escalations than do smaller projects; for rail and road projects this does not appear to be the case.

- For all project types, our data do not support that bigger projects have a larger risk of cost escalation than do smaller ones; the risk of cost escalation is high for all project sizes and types.

- Projects are growing larger over time, but only significantly so for road projects.

Because the same percentage cost escalation will typically cause more problems in a large project than in a small one, we conclude that an increase in project size translates into a need for improved planning processes and institutional set-ups for infrastructure development and management.

Third, for type of ownership we find that the data do not support the often seen claim that public ownership is problematic per se and private ownership a main source of efficiency in curbing cost escalation. This, however, does not rule out the possibility that other reasons may exist for preferring private over public ownership; for instance, that private ownership may help protect the ordinary taxpayer from financial risk and may reduce the number of people exposed to such risk. The data show, nevertheless, that the issue of ownership is more complex than is usually assumed. The main problem in relation to cost escalation may not be public versus private ownership but a certain kind of public ownership, namely state-owned enterprises, which lack *both* the transparency and public control that placement in the public sector proper would entail *and* the competitive pressure that placement in the private sector would bring about. We expect further research on this issue to be particularly rewarding in either falsifying or confirming this finding. It is an issue of principal significance for deciding on the institutional set-up and regulatory regime for infrastructure provision.

*Table 1: Average cost escalation and ownership for fixed links. 15 projects, constant prices.*

| Ownership | Number of cases (N) | Average cost escalation | Standard deviation |
|---|---|---|---|
| Private | 4 | 34.0 | 30.1 |
| State-owned enterprise | 3 | 110.0 | 71.5 |
| Other public | 8 | 23.1 | 33.6 |

*Table 2: Ownership and percentage cost escalation in 25 rail projects. Constant prices.*

| Ownership | No. of projects (N) | High-speed rail | Urban rail | Conventional rail |
|---|---|---|---|---|
| State-owned enterprise | 9 | 88.0 | 35.5 | - |
| Other public ownership | 16 | 15.0 | 53.5 | 29.6 |

Figure 1: Length of implementation phase and cost escalation in 111 transport infrastructure projects, constant prices. For the regression line, the ten projects with implementation phases of 13 years or longer are considered as outliers.

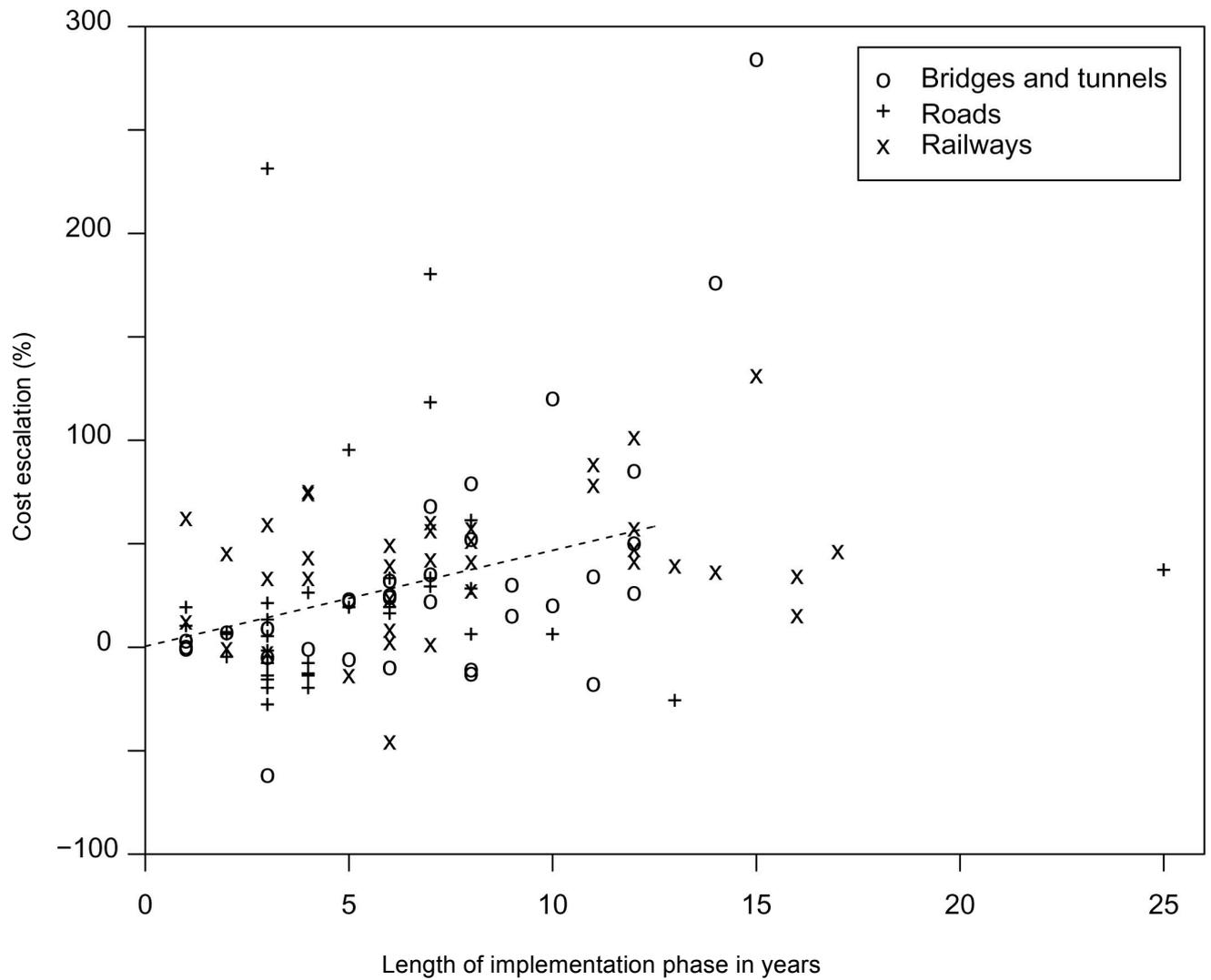

Figure 2: Box plots of length of implementation phase.

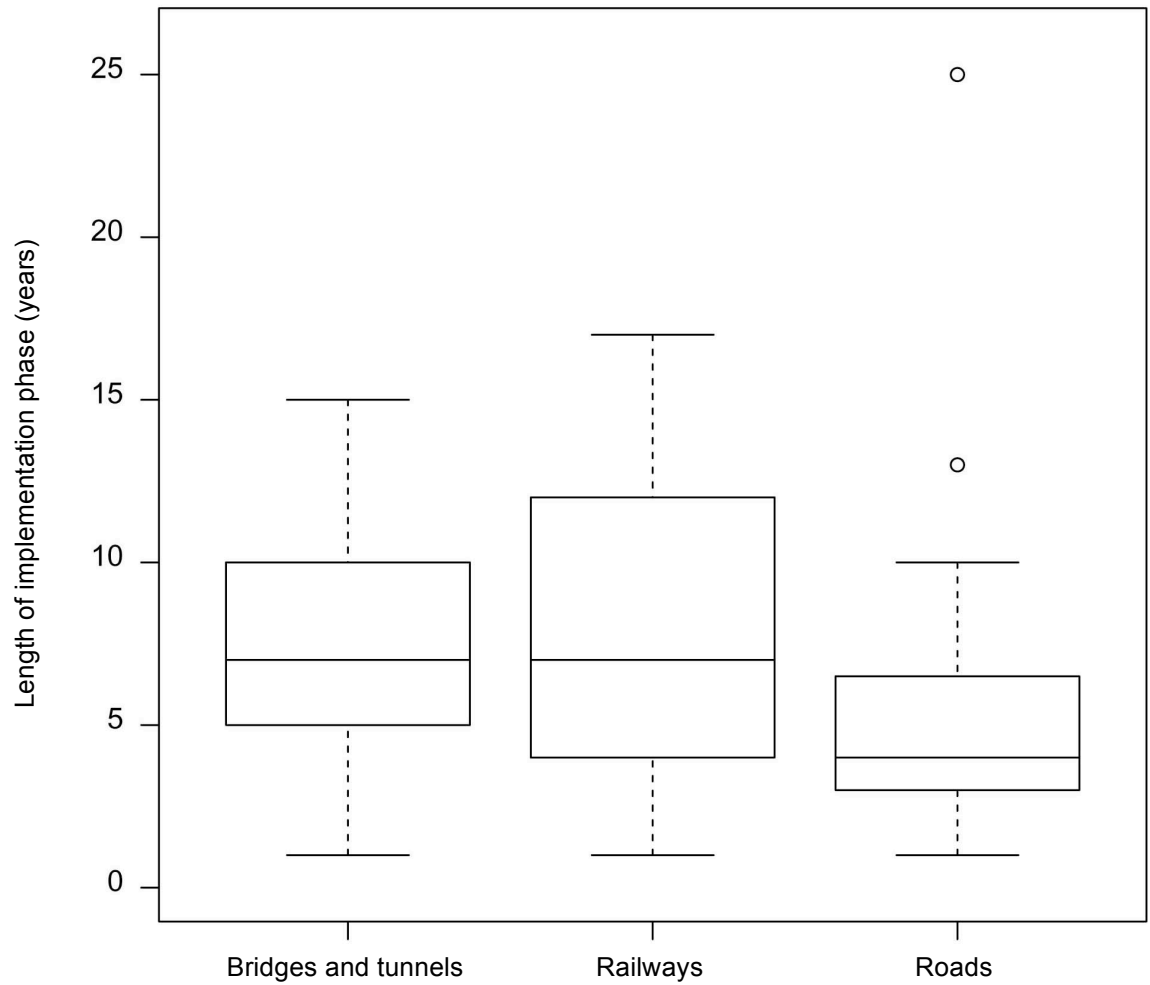

Figure 3: Forecast construction costs and cost escalation in 131
transport infrastructure projects, constant 1995 prices (EUR 1 = US$
1.29; 1995).

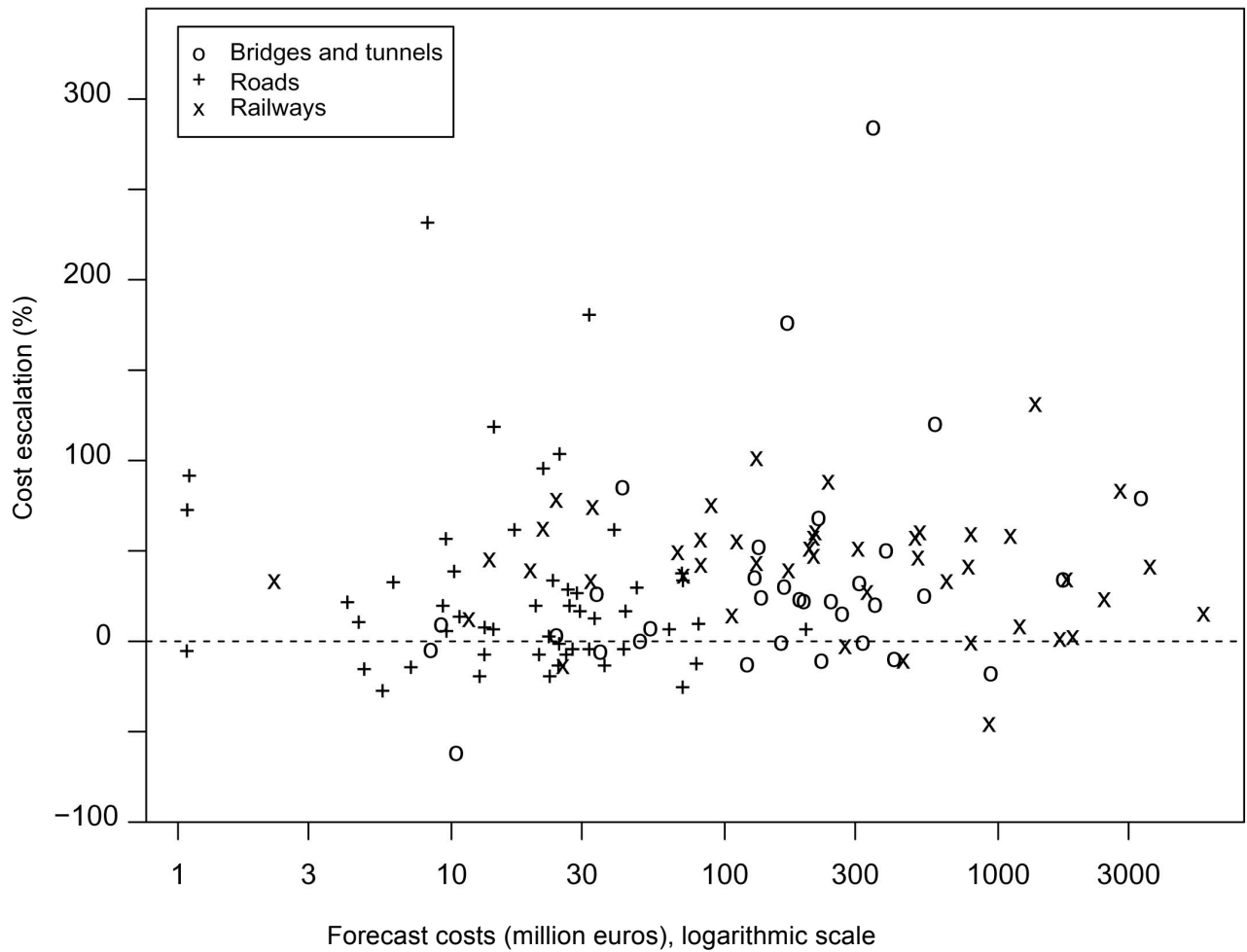

Figure 4: Size of projects 1925-2000. Year of completion and actual construction costs. Constant 1995 prices, logarithmic scale. 131 projects. (M=million euros, B=billion euros; EUR 1 = US$ 1.29; 1995).

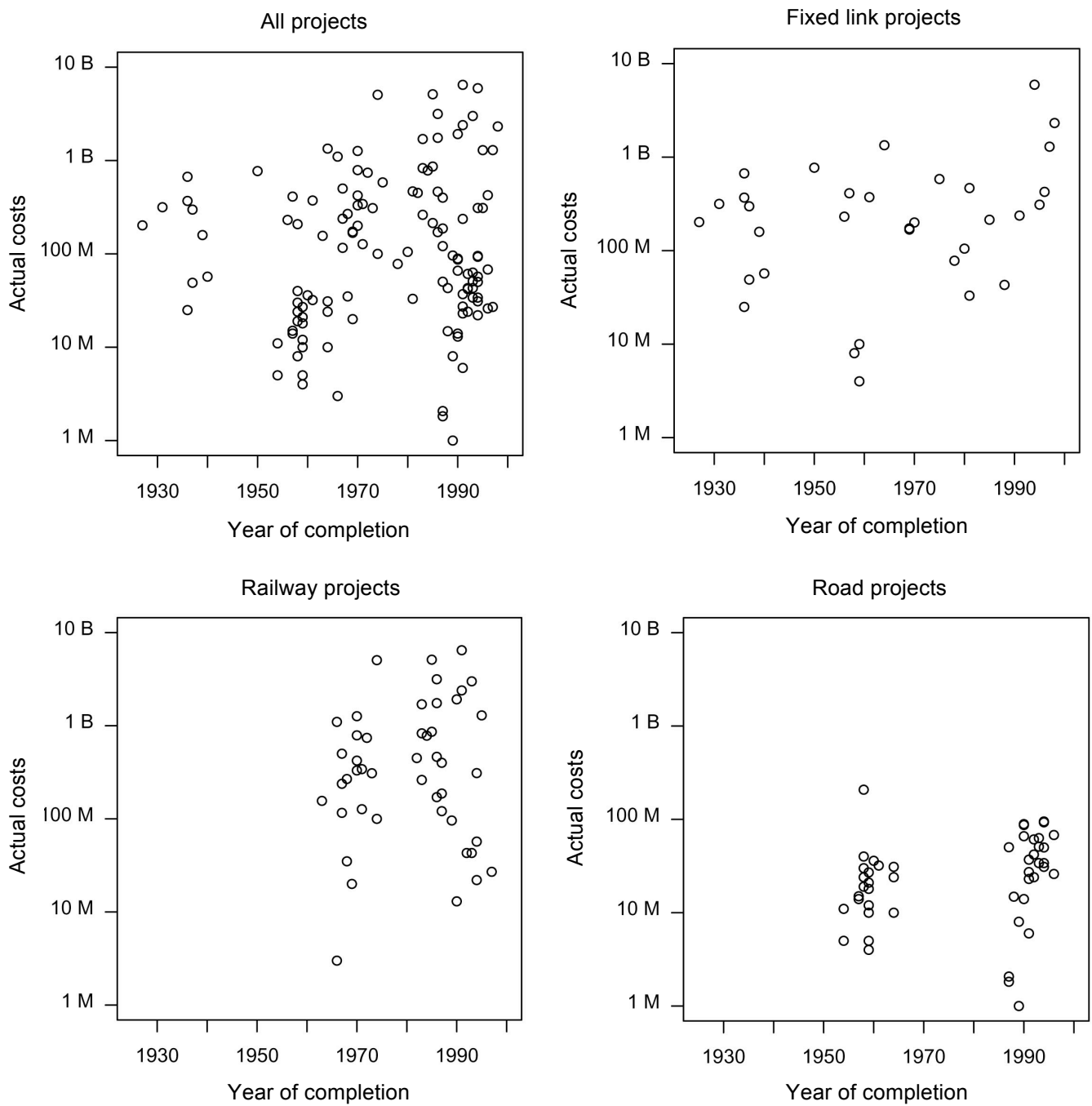